\documentclass[aps,prd,twocolumn,superscriptaddress,nofootinbib,preprintnumbers,10pt]{revtex4-2}
\usepackage{style}

\begin{document}

\title{The NANOGrav 15-year data set: A Limit on Continuous Spin Gravitons}

\author{Chinmay Ambasht\,\orcidlink{0009-0005-3928-3807}}
\email{ambasht2@illinois.edu}
\affiliation{Department of Physics, Grainger College of Engineering, University of Illinois Urbana-Champaign, Urbana, IL 61801, USA}

\author{Olivia Bitcon\,\orcidlink{0000-0002-2962-2191}}
\email{obitcon2@illinois.edu}
\affiliation{Department of Physics, Grainger College of Engineering, University of Illinois Urbana-Champaign, Urbana, IL 61801, USA}

\author{Tanner Trickle\,\orcidlink{0000-0003-1371-4988}}
\email{ttrickle@illinois.edu}
\affiliation{Department of Physics, Grainger College of Engineering, University of Illinois Urbana-Champaign, Urbana, IL 61801, USA}

\author{The NANOGrav Collaboration}
\thanks{See App.~\ref{app:complete_author_list} for the NANOGrav Collaboration author list.}

\begin{abstract}
Continuous spin particles (CSPs) are massless particles characterized by a dimensionful spin scale, $\rho$, which, when non-zero, changes their helicity under Lorentz boosts. Since CSPs reduce to ordinary particles when $\rho \rightarrow 0$, the photon or graviton could be CSPs and only appear to behave as ordinary particles due to a small value of $\rho$. Therefore, it is important to determine what value of $\rho$ is experimentally allowed. Recent work has shown that the physical effects of CSPs scale as powers of $\rho/\omega$ when $\rho \ll \omega$, where $\omega$ is a frequency scale, making low-frequency probes particularly beneficial for highlighting these phenomena. In this paper we use pulsar timing arrays to set constraints on the spin scale of the continuous spin graviton. Pulsar timing arrays are uniquely suited for this application as they have exceptional sensitivity to low, nanohertz frequency gravitational waves, as demonstrated by their recent evidence for a stochastic gravitational wave background. Using standard Bayesian data analysis techniques, we find that the NANOGrav 15-year data set constrains the graviton spin scale to $\rho_\text{g} < 1.23 \times 10^{-23} \, \text{eV} $ at the 95\% credible level.
\end{abstract}

\maketitle

\tableofcontents

\section{Introduction}
\label{sec:introduction}

Recent advances in observational techniques have opened a new era of precision gravitational wave (GW) astronomy.
While modern interferometers continually provide direct measurements of GWs~\cite{LIGOScientific:2016aoc,Cahillane:2022pqm}, pulsars provided the first indirect evidence for their existence~\cite{Hulse:1974eb,Taylor:1982zz}.
Recently, pulsar timing arrays (PTAs), including the Chinese PTA (CPTA)~\cite{Xu:2023wog}, European PTA (EPTA) and Indian PTA (InPTA)~\cite{EPTA:2023fyk}, NANOGrav~\cite{NANOGrav:2023gor}, and the Parkes PTA (PPTA)~\cite{Reardon:2023gzh}, saw the first evidence for a stochastic GW background (SGWB)~\cite{InternationalPulsarTimingArray:2023mzf}.
These precise GW detectors can also be used to search for physics beyond the Standard Model from, for example, novel cosmological sources of SGWBs~\cite{Vilenkin:1984ib,Guzzetti:2016mkm,Caprini:2019egz,Domenech:2021ztg,NANOGrav:2023hvm,Agazie:2024RSI,Agazie:2026PPL}, astrophysical imprints of ultra-heavy~\cite{Dror:2019twh,Foster:2026kfg,Cherukupalli:2026cda} or ultra-light dark matter~\cite{Khmelnitsky:2013lxt,EuropeanPulsarTimingArray:2023egv,Foster:2026mvs}, or modifications to gravity arising from a massive graviton, additional polarization states, or otherwise modified graviton dispersion relations~\cite{Lee:2010cg,Lee:2013awh,Gair:2015hra,Qin:2020hfy,Liang:2023ary}.

In this paper we investigate the physical effects a ``continuous spin" graviton (CSG) would have on the observed timing residuals in PTAs.
Continuous spin particles are ordinary massless particles in the sense that they are irreducible unitary representations of the Poincar\'e group~\cite{Schuster:2013pxj,Schuster:2013vpr,Schuster:2014hca,Rivelles:2014fsa,Rivelles:2016rwo,Bekaert:2017khg,Najafizadeh:2017tin,Schuster:2023xqa,Bellazzini:2024dco}. 
However, these continuous spin particles are very different from ordinary massless particles, which are typically assumed to be null eigenstates of $W^2 = W^\mu W_\mu$, where $W^\mu = \epsilon^{\mu \nu \rho \sigma} J_{\nu \rho} P_\sigma / 2$ is the Pauli-Luba\'nski vector, $J_{\mu \nu}$ is the angular momentum, and $P_\mu$ is the four-momentum.
This restriction limits the number of helicities to the usual two and forces the helicities to be Lorentz invariant.
This restriction can be loosened by considering particles parameterized by a dimensionful spin scale, $\rho$, which satisfies $W^2 = - \rho^2$~\cite{Schuster:2013pxj,Schuster:2023xqa}.
These continuous spin particles are composed of more than two helicity modes and smoothly reduce to ordinary particles in the $\rho \rightarrow 0$ limit.
Furthermore, they evade the Weinberg-Witten theorems~\cite{Weinberg:1964ev,Weinberg:1964ew,Weinberg:1965rz,Weinberg:1980kq} limiting the long-range phenomenology of particles with helicity greater than two, since the helicity is not Lorentz invariant when $\rho \neq 0$.

While the existence of continuous spin particles has been known since Wigner's original classification~\cite{Wigner:1939cj}, it was only recently understood how to write down a Lagrangian for continuous spin fields that couple to the Standard Model~\cite{Schuster:2023xqa}.
Given this consistent description for continuous spin particles that can interact with matter, it is possible that the massless particles in the Standard Model, e.g., photons and gravitons, could be continuous spin particles with small values of $\rho$. It is therefore an important question to understand the value of $\rho$ that is experimentally allowed for each massless particle.

The interferometric phenomenology of a CSG was recently derived in Ref.~\cite{Kundu:2025fsd}, and we follow their methodology closely to understand the effect the CSGs' spin scale, $\rhog$, has on the timing residuals in the NANOGrav 15-year data set~\cite{NANOGrav:2023hde}.
As in Ref.~\cite{Kundu:2025fsd}, we assume that the photon's spin scale is zero.
Ref.~\cite{Kundu:2025fsd} showed that the effects of a CSG on the timing delay induced by a monochromatic GW appear at $\mathcal{O}(\rhog^2 / \omega^2)$, for small $\rhog / \omega$, where $\omega$ is the frequency of the GW. Therefore PTAs are a natural probe of small $\rhog$, since they probe the lowest frequency GWs that we have evidence for.
Given this, the preliminary expectation is that PTAs are sensitive to $\rhog \sim 2 \pi / T \sim 10^{-23} \, \text{eV} \left( 15 \, \text{yr} / T \right)$, where $T$ is the observation time of the PTA, normalized to 15 years since that is the observation time of the NANOGrav data set we consider here.
We formalize this expectation by performing a complete Bayesian analysis searching for the effects of a CSG in the NANOGrav 15-year data set, and use this to set an upper limit on the allowed value of $\rhog$.

This paper is organized as follows.
In Sec.~\ref{sec:theory} we derive the CSG-induced, $\rhog$-dependent, timing residual in the context of PTAs. We closely follow the formalism developed for interferometric observables in Ref.~\cite{Kundu:2025fsd}, and emphasize the modifications to the standard Hellings-Downs (HD) overlap reduction function (ORF) (Sec.~\ref{subsec:modified_orf}).
In Sec.~\ref{sec:data_analysis} we discuss the data analysis techniques used to constrain $\rhog$ with the NANOGrav 15-year data set. We begin with a discussion of the modified, $\rhog$-dependent, marginalized likelihood (Sec.~\ref{subsec:likelihood}), and then summarize our Bayesian inference pipeline, which relies on \textsf{PTArcade}~\cite{Mitridate:2023oar} (Sec.~\ref{subsec:bayesian_inference}).
Lastly, in Sec.~\ref{sec:sensitivity} we illustrate the sensitivity of the NANOGrav 15-year data set to $\rhog$, finding the 95\% credible level limit to be $\rhog < 1.23 \times 10^{-23} \, \text{eV}$.
Throughout we work in natural units ($\hbar = c = 1$), use a mostly minus Minkowski metric signature, $\eta^{\mu \nu} \sim \text{diag}(+, -, -, -)$, and use $a \cdot b$ to denote contraction with the Minkowski metric: $a \cdot b \equiv \eta_{\mu \nu}a^\mu b^\nu$.

\begin{figure*}[ht!]
    \centering
    \includegraphics[width = 0.8\paperwidth]{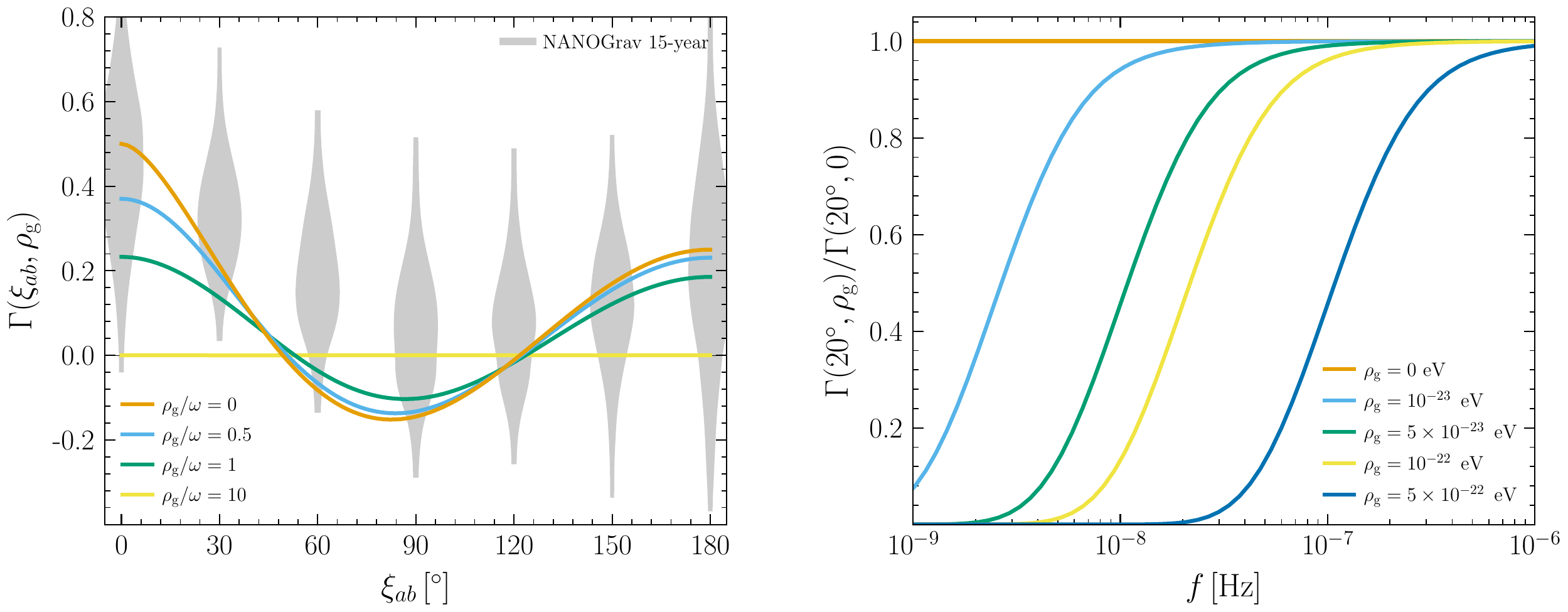}
    \caption{Dependence of the modified overlap reduction function (ORF), $\Gamma(\xi_{ab})$ (Eq.~\eqref{eq:money_ORF}), on the continuous spin scale of the graviton, $\rhog$. The orange line in both plots represents the Hellings-Downs (HD) ORF, corresponding to the $\rhog = 0$ limit of Eq.~\eqref{eq:money_ORF}. \textbf{Left}. The modified ORF as a function of the pulsar separation angle, $\xi_{ab}$, for various values of $\rho_\text{g}/\omega$. The measured posterior densities of the HD ORF from the NANOGrav 15-year data set are shown as gray violins~\cite{NANOGrav:2023gor}. As $\rhog / \omega$ increases, the angular correlations in the ORF dampen with a slight shift in the roots of ORF. \textbf{Right}. Ratio of the modified ORF to the HD ORF as a function of frequency, for a fixed pulsar separation angle of $\xi_{ab} = 20^\circ$. We see that for $f \lesssim \rhog / 2 \pi$ the modified ORF is damped in frequency space relative to the HD ORF.}
    \label{fig:modified_orf}
\end{figure*}

\section{Continuous Spin Graviton Induced Timing Residual}
\label{sec:theory}

PTAs are exceptionally sensitive to changes in the arrival time of light emitted by pulsars.
Once ephemeris effects~\cite{2021AJ....161..105P} are accounted for, the arrival time of a light pulse that was emitted at time $t_\text{em}$ can be modeled as,
\begin{align}
    t_\text{arr} = t_\text{em} + d + \delta t \, ,
    \label{eq:arrival_time}
\end{align}
where $\mathbf{d}$ is the position vector pointing from the Earth to the pulsar, $d = |\mathbf{d}|$, $\mathbf{d} = d\,\hat{\mathbf{d}}$, and $\delta t$ is the change in arrival time caused by the CSG.\footnote{The arrival time of the $n^\text{th}$ pulse is found by substituting the emitted time of the $n^\text{th}$ pulse into Eq.~\eqref{eq:arrival_time}, which is calculated as $\phi(t_\text{em}^n) = 2 \pi n$, where $\phi(t)$ is the phase evolution of the pulsar.}
The purpose of this section is to derive $\delta t$ at linear order in the GWB amplitude, starting with monochromatic GWs in Sec.~\ref{subsec:mono}, and then focusing on a SGWB in Sec.~\ref{subsec:SGWB}.

To calculate the photon arrival time we must calculate the photon trajectory, $x^\mu$, which we parameterize by $\lambda$; at $\lambda = 0$ the light is emitted from the pulsar, $x^\mu(\lambda = 0) = (t_\text{em}, \mathbf{d})$, and at $\lambda = 1$ the light arrives at Earth, $x^\mu(\lambda = 1) = (t_\text{arr}, 0)$.
Since our goal is to calculate $\delta t$ to linear order in the GWB amplitude, we expand $x^\mu$ perturbatively as $x^\mu = x_0^\mu + \delta x^\mu$, where $x_0^\mu$ is the trajectory in the absence of the GWB, and $\delta x^\mu$ is linear in the GWB amplitude.
The zeroth-order trajectory is simply,
\begin{align}
    x_0^\mu & = (\lambda d + t_\text{em},  (1 - \lambda) \, \mathbf{d})~,~ \frac{dx^\mu_0}{d\lambda} = (d, -\mathbf{d}) \, .
    \label{eq:zeroth_order}
\end{align}
Furthermore, we can directly relate the change in arrival time to the change in photon trajectory since $\delta x^\mu(\lambda = 1) = (\delta t, 0)$, and therefore,
\begin{align}
    \delta t = \frac{1}{d}\frac{dx_0 }{d\lambda}\cdot \delta x\bigg|_{\lambda = 1} = \frac{1}{d} \int_0^1  \frac{dx_0}{d\lambda} \cdot \frac{d \hspace{0.1em} \delta x}{d\lambda} \; \text{d} \lambda \, .
    \label{eq:delta_t_connection}
\end{align}
This specific parameterization has been chosen since $(dx_0/d\lambda) \cdot ( d \hspace{0.1em} \delta x / d\lambda)$ is the quantity that naturally arises when solving for $\delta x$, as discussed in detail next.

\subsection{Monochromatic Gravitational Wave}
\label{subsec:mono}

The derivation leading up to Eq.~\eqref{eq:delta_t_connection} uses the same procedure one could use to calculate $\delta t$ in an ordinary GWB~\cite{Maggiore:2007ulw,Taylor:2021yjx}.
The CSG spin scale modifies this calculation by changing the photon Lagrangian, and therefore changing $x^\mu$, in the presence of a GWB~\cite{Kundu:2025fsd}.
Assuming a monochromatic GW with amplitude $h_A$ in the $A^{\text{th}}$ polarization, $A \in \{ +, \times \}$, and four-momentum $k^\mu = \omega \, (1, \hat{\mathbf{\Omega}})$, where $\omega$ is the GW frequency and $\hat{\mathbf{\Omega}}$ is its direction, the $\rhog$-dependent photon Lagrangian to linear order in $h_A$ is, in the worldline formalism~\cite{Kundu:2025fsd},
\begin{align}
     L = & \frac{1}{e(\lambda)} \bigg[ \frac{1}{d^2} \frac{dx_0}{d\lambda} \cdot \frac{d \hspace{0.1em} \delta x}{d\lambda} \nonumber \\ 
     & + (1 + \hat{\mathbf{\Omega}} \cdot \hat{\mathbf{d}}) \, \left( \frac{8 J_2(\bar{\rho})}{\bar{\rho}^2} \right) \sum_A F_A \, h_A  \cos(k \cdot x_0) \bigg] \, ,
     \label{eq:modified_photon_L}
\end{align}
where $e(\lambda)$ is the einbein function, we have assumed for simplicity that the GW phase is zero for each polarization, and have defined both a dimensionless spin scale parameter,
\begin{equation} 
    \bar{\rho} \equiv \frac{\rhog}{\omega} \sqrt{\frac{1 - \hat{\bm{\Omega}} \cdot \hat{\mathbf{d}} }{1 + \hat{\bm{\Omega}} \cdot \hat{\mathbf{d}}} } ~,
    \label{eq:dimensionless_spin_scale}
\end{equation}
and the antenna function, $F_A$,~\cite{Taylor:2021yjx}
\begin{align}
    F_A \equiv \frac{1}{2} \frac{\hat{\mathbf{d}}^i \hat{\mathbf{d}}^j }{1 + \hat{\mathbf{\Omega}} \cdot \hat{\mathbf{d}}} e^{A}_{ij} \, ,
    \label{eq:geometric_factor}
\end{align}
where $e_A^{ij}$ is the $A^{\text{th}}$ GW polarization tensor in the transverse-traceless frame.
Furthermore, we note that in the $\rhog \rightarrow 0$ limit, $8 J_2(\bar{\rho}) / \bar{\rho}^2 \rightarrow 1$, which is a useful simplification when comparing to the standard result in general relativity.

To calculate $\delta x^\mu$, one could use the Euler-Lagrange equations from the Lagrangian in Eq.~\eqref{eq:modified_photon_L}. However the einbein constraint, which simply sets the term in the block parentheses to zero, is much more useful here since it directly calculates $(dx_0 / d\lambda) \cdot (d\hspace{0.1em}\delta x / d \lambda)$, which is all that is needed to find the change in arrival time using Eq.~\eqref{eq:delta_t_connection}.
Following this procedure, the final CSG-induced timing residual is,
\begin{align}
    \delta t =  \frac{1}{\omega} & \frac{ 8J_2(\bar{\rho})}{\bar{\rho}^2} \sum_A F_A \, h_A \nonumber\\
    & \times \bigg[ \sin \left( \omega (t - d) - \omega \hat{\mathbf{\Omega}} \cdot \mathbf{{d}} \right) - \sin \left( \omega t\right)  \bigg] \, ,
    \label{eq:CSG_timing residual}
\end{align}
which we have written in terms of the zeroth-order arrival time instead of the emitted time, $t \equiv t_\text{arr} \approx t_\text{em} + d$, since the data set is defined in terms of the photon arrival times.
The two terms in Eq.~\eqref{eq:CSG_timing residual} have the familiar interpretation of the ``pulsar" and ``Earth" terms, respectively.

\subsection{Stochastic Gravitational Wave Background}
\label{subsec:SGWB}

We now discuss the CSG-induced timing residual in the presence of a SGWB, whose metric fluctuations in the $\rhog = 0$ limit are defined as,
\begin{align}
    h_{\mu \nu}(\mathbf{x}, t) = \sum_A\int \dd f \, \dd^2 \hat{\mathbf{\Omega}} \, e^{-i \omega t} e^{i \omega \hat{\mathbf{\Omega}} \cdot \mathbf{x} }\, h_A(f, \hat{\mathbf{\Omega}}) \, e^A_{\mu \nu} \, ,
\end{align}
where $\omega = 2 \pi f$, $h_A(f, \hat{\mathbf{\Omega}})$ are the mode amplitudes satisfying $h_A^*(f, \hat{\mathbf{\Omega}}) = h_A(-f, \hat{\mathbf{\Omega}})$, and the $f$ integral is over positive and negative frequencies.
Here we calculate the CSG-induced timing residual as in Sec.~\ref{subsec:mono}, but with a summation over all possible mode amplitudes.
The final timing residual is,
\begin{align}
    \delta t_\text{SGWB} = \int \dd f \, \dd^2& \hat{\mathbf{\Omega}} \frac{i}{\omega} \frac{8J_2(\bar{\rho})}{\bar{\rho}^2} \sum_A F_A \, h_A(f, \hat{\mathbf{\Omega}}) \nonumber \\
    & \times \left[ e^{- i \omega t} \left( e^{i\omega d (1 + \hat{\mathbf{\Omega}} \cdot \hat{\mathbf{d}}) }  - 1\right) \right] \, ,
    \label{eq:CSG_SGWBtiming residual}
\end{align}
which is the generalization of Eq.~\eqref{eq:CSG_timing residual} expected from linearity.

\subsection{Modified Overlap Reduction Function}
\label{subsec:modified_orf}

The ORF, $\Gamma(\xi_{ab})$, determines the timing residual correlation pattern in a PTA as a function of the separation angle, $\xi_{ab}$, between pulsars $a$ and $b$, defined by $\cos(\xi_{ab}) \equiv \hat{\mathbf{d}}_a \cdot \hat{\mathbf{d}}_b$.
In the $\rhog = 0$ limit, the ORF is the HD curve~\cite{Hellings:1983fr}.
When $\rhog \neq 0$, the ORF is modified due to the timing residual changes in Eq.~\eqref{eq:CSG_SGWBtiming residual}.
Assuming that the SGWB is Gaussian-stationary, unpolarized, spatially homogeneous and isotropic, the GWB mode amplitudes, $h_A(f, \hat{\mathbf{\Omega}})$, are defined by their two-point statistics~\cite{Taylor:2021yjx},
\begin{align}
    \langle h_A(f, \hat{\mathbf{\Omega}})h^*_{A'}(f', \hat{\mathbf{\Omega}}')\rangle = \frac{\delta_{AA'}\delta^2(\hat{\mathbf{\Omega}} - \hat{\mathbf{\Omega}}')\delta(f - f')}{16 \pi} S_h(f) \, ,
    \label{eq:mode_correlations}
\end{align}
where $S_h(f)$ is the one-sided power spectral density.
We additionally assume that the produced SGWB is unaffected by a finite value of $\rhog$, and discuss this assumption further in Sec.~\ref{sec:conclusions}.

The modified ORF is then calculated by substituting Eq.~\eqref{eq:CSG_SGWBtiming residual} in to the definition of the ORF~\cite{Taylor:2021yjx},

\begin{align}
    \langle \dot{\delta t}_a(f) \dot{\delta t}_b^*(f') \rangle = \frac{1}{3}\Gamma(\xi_{ab}) \,  \frac{\delta(f - f')}{2} \,  \, S_h(f) \, ,
    \label{eq:defining_ORF}
\end{align}

\noindent where $\dot{\delta t}_a$ is the time derivative of the timing residuals of pulsar $a$, $\dot{\delta t}_a(f) \equiv \int \dd t \, e^{2 \pi i f t} \, \dot{\delta t}(t)$ is its Fourier transform, and Eq.~\eqref{eq:mode_correlations} is used to compute the expectation value in Eq.~\eqref{eq:defining_ORF}.
The resulting modified ORF for $a \neq b$ is given by,
\begin{widetext}
\begin{align}
    \Gamma(\xi_{ab}) = \frac{3}{8} \int \frac{\text{d}^2 \hat{\mathbf{\Omega}}}{4 \pi}  & \left[ \frac{8 J_2(\bar{\rho}_a)}{\bar{\rho}_a^2} \right] \left[ \frac{8 J_2(\bar{\rho}_b)}{\bar{\rho}_b^2} \right] \frac{1}{(1 + \hat{\mathbf{\Omega}} \cdot \hat{\mathbf{d}}_a)(1 + \hat{\mathbf{\Omega}} \cdot \hat{\mathbf{d}}_b)} \nonumber \\
    & \times \bigg\{ 2 \left[ \hat{\mathbf{d}}_a \cdot  \hat{\mathbf{d}}_b  - \left( \hat{\mathbf{\Omega}} \cdot \hat{\mathbf{d}}_a\right) \left( \hat{\mathbf{\Omega}} \cdot \hat{\mathbf{d}}_b\right) \right] ^2 
    - \bigg[ 1 - \left( \hat{\mathbf{\Omega}} \cdot \hat{\mathbf{d}}_a\right)^2\bigg] \left[ 1 - \left(\hat{\mathbf{\Omega}} \cdot \hat{\mathbf{d}}_b\right)^2 \right] \bigg\} \, ,
    \label{eq:money_ORF}
\end{align}
\end{widetext}
where $\bar{\rho}_a$ is $\bar{\rho}$ defined in Eq.~\eqref{eq:dimensionless_spin_scale} evaluated for pulsar $a$. When $a = b$ the ORF in Eq.~\eqref{eq:money_ORF} should be multiplied by $2$~\cite{Taylor:2021yjx}. Additionally, note that while Eq.~\eqref{eq:money_ORF} does not immediately appear to depend on only $\xi_{ab}$, one can rotate to the ``computational frame"~\cite{Mingarelli:2013dsa}, where,
\begin{align}
    \hat{\mathbf{d}}_a = (0, 0, 1)~~~,~~~\hat{\mathbf{d}}_b &= (\sin\xi_{ab}, 0, \cos \xi_{ab}) \, ,
    \label{eq:computational_frame}
\end{align}
rendering the ORF dependent on just $\xi_{ab}$. In the $\rhog = 0$ limit, Eq.~\eqref{eq:money_ORF} reduces to the HD curve.

In Fig.~\ref{fig:modified_orf} we show how the ORF is modified when $\rhog \neq 0$.
Since the effects of $\rhog$ enter as $\rhog / \omega$ in Eq.~\eqref{eq:money_ORF}, there are two primary effects associated with a non-zero CSG spin-scale.
First, when $\rhog / \omega \gg 1$ the correlation structure is damped at all angles. This can be seen in the left panel of Fig.~\ref{fig:modified_orf}.
Second, for fixed $\rhog$ there is a frequency cutoff in the modified ORF at $f \sim \rhog / 2 \pi$, which can be seen in the right panel of Fig.~\ref{fig:modified_orf}.
These changes in the angular and frequency dependence of the ORF relative to the HD curve are the defining features of the CSG-modified ORF.
Details of the numerical evaluation of the CSG-modified ORF (Eq.~\eqref{eq:money_ORF}) are given in App.~\ref{app:modified_orf_eval}. 

\section{Data Analysis}
\label{sec:data_analysis}

In this section we discuss how to search for CSGs in PTA datasets using the standard Bayesian analysis framework.
In Sec.~\ref{subsec:likelihood}, we discuss the timing residual model and marginalized likelihood to highlight where the CSG modifies the standard derivation.
In Sec.~\ref{subsec:bayesian_inference}, we discuss our Bayesian inference pipeline which computes the posterior distribution of $\rho_\text{g}$.

\subsection{Timing Residuals and Marginalized Likelihood}
\label{subsec:likelihood}

\begin{table*}[ht!]
    \begin{center}
        \renewcommand{\arraystretch}{1.3}
        \setlength{\tabcolsep}{6.5pt}
        \small
            \begin{tabular}{llll}
                \toprule
                    \multicolumn{1}{c}{\bf{Parameter}}  & \multicolumn{1}{c}{\bf{Description}} & \multicolumn{1}{c}{\bf{Prior}} & \multicolumn{1}{c}{\bf{Comments}} \\
                \midrule
                    \multicolumn{4}{c}{\bf{White Noise}} \\[1pt]
                    $E_{b}$ & EFAC per backend/receiver system & Uniform $[0.1, 5]$ & one parameter per pulsar \\
                    $Q_{b}$ [s] & EQUAD per backend/receiver system & log-Uniform $[-8.5, -5]$ & one parameter per pulsar  \\
                \midrule
                    \multicolumn{4}{c}{\bf{Red Noise}} \\[1pt]
                    $A_{{\rm red}, b}$ & red noise power-law amplitude & log-Uniform $[-20, -11]$ & one parameter per pulsar  \\
                    $\gamma_{{\rm red}, b}$ & red noise power-law spectral index & Uniform $[0, 7]$ & one parameter per pulsar \\
                \midrule
                    \multicolumn{4}{c}{\bf{Supermassive Black Hole Binaries (SMBHB)}} \\[1pt]
                    $A_{\mathrm{GWB}}$ & common process strain amplitude & log-Uniform $[-20, -11]$ & one parameter for PTA \\
                    $\gamma_{\mathrm{GWB}}$ & common process power-law spectral index & Uniform $[0, 7]$ & one parameter for PTA \\
                \midrule
                    \multicolumn{4}{c}{\bf{Continuous Spin Graviton Spin Scale}} \\[1pt]
                    $\log_{10}(\rho_\text{g} / \mathrm{eV})$ & dimensionless CSG parameter & log-Uniform $[-30.18, -20.98]$
 & one parameter for PTA \\
                \bottomrule
            \end{tabular}
        \caption{Prior distributions used in the Bayesian data analysis discussed in Sec.~\ref{subsec:bayesian_inference}.}
        \label{tab:parameters}
    \end{center}
\end{table*}

We begin with a discussion of the timing residuals and marginalized likelihood used in our analysis. Our methodology follows standard procedures when searching for new physics in PTA datasets; see Refs.~\cite{Kaplan:2022lmz,NANOGrav:2023hvm} for more detailed discussions. The purpose of reviewing the formalism here is to precisely describe where the CSG spin scale, $\rhog$, appears, and to define the other timing model parameters we use, which are summarized in Table~\ref{tab:parameters}.

Following a preliminary fit of the timing-ephemeris parameters to the TOAs to account for known astrophysical effects, the $i^\text{th}$ timing residual at the $b^\text{th}$ pulsar is modeled as,
\begin{align}
    \Delta t_{ib} = n_{ib} + F_{i\alpha}\,a_{\alpha b} + M_{i m} \epsilon_{m b} \, ,
    \label{eq:timing_model}
\end{align}
where repeated indices are implicitly summed over, $F_{i \alpha}$ is the Fourier design matrix, whose odd (even) columns are simply $F_{i \alpha} = \sin(2\pi i  \alpha / T)$ ($F_{i \alpha} = \cos(2\pi i \alpha / T)$), $T$ is the observation time, and $M_{im}$ is the design matrix.
Each term in Eq.~\eqref{eq:timing_model} has a simple interpretation: the first term is associated with white noise, the second term is from chromatic (e.g., red) noise, and the last term allows for small deviations in each of the timing-ephemeris parameters.
$n_{ib}, a_{\alpha b}$ and $\epsilon_{mb}$ are all zero-mean Gaussian random variables, with covariances governed by the timing model parameters, which we will now discuss in detail.

The white noise covariance is,
\begin{align}
    \langle n_{ib} \,n_{j c} \rangle & = E_b^2 \, \sigma_i \, \delta_{ij} \, \delta_{bc} + Q_b^2 \, \delta_{ij} \, \delta_{bc} \label{eq:white_noise_parameterization} \, ,
\end{align}
where $E_b$ and $Q_b$ are the ``EFAC" and ``EQUAD" white noise parameters, respectively, and $\sigma_i$ is the time-of-arrival uncertainty for the $i^\text{th}$ observation. The red noise covariance contains two sources, one from the stochastic GWB (nominally coming from the inspiral of supermassive black hole binaries (SMBHB)), labeled ``GWB," and the other from intrinsic pulsar red noise, labeled ``red,"
\begin{align}
    \langle a_{\alpha b} \, a_{\beta c} \rangle & = \Gamma_{bc}(\rho_\text{g}) \,  \, \Phi^\text{GWB}_\alpha \, \delta_{\alpha \beta} + \Phi_{\alpha b}^\text{red}  \, \delta_{\alpha \beta} \, \delta_{bc} \, ,
    \label{eq:fourier_cov}
\end{align}
where $\Gamma_{bc}$ is the ORF, that depends on the CSG spin scale, $\rho_\text{g}$, as discussed in Sec.~\ref{sec:theory} and given in Eq.~\eqref{eq:money_ORF}. In the $\rho_\text{g} \rightarrow 0$ limit, $\Gamma_{bc}$ reduces to the standard HD correlation function.
$\Phi^\text{GW}_\alpha = \Phi^\text{GW}(f = \alpha / T)$, $\Phi^\text{red}_{\alpha b} = \Phi^\text{red}_b(f = \alpha / T)$, and $\Phi^\text{GWB}(f), \Phi^\text{red}_b(f)$ describe the spectral features of these red noises and are parameterized by,
\begin{align}
    \Phi^\text{GWB} & = \frac{H_0^2 \, \Omega_\text{GWB}}{8 \pi^4 f^5 \, T} = \frac{h^2 A_\text{GWB}^2}{12 \pi^2 \, T f_\text{ref}^3}  \left( \frac{f}{f_\text{ref}} \right)^{-\gamma_\text{GWB}} \\
    \Phi^\text{red}_b & = \frac{h^2 A_{\text{red},b}^2}{12 \pi^2 \, T f_\text{ref}^3}  \left( \frac{f}{f_\text{ref}} \right)^{-\gamma_{\text{red}, b}} \, ,
\end{align}
where $H_0 \approx 70 \, \text{km} / \text{s} / \text{Mpc}$ is the present day Hubble constant, $h \approx 0.7$ is the reduced Hubble constant, $\rho_c \Omega_\text{GWB} = d \rho_\text{GWB} / d \ln f$, where $\rho_c$ is the critical density, $\rho_\text{GWB}$ is the GWB energy density, and $f_\text{ref}$ is a reference frequency, taken to be $f_\text{ref} = 1 / \text{yr}$. $A_\text{GWB}, \gamma_\text{GWB}, A_{\text{red}, b}$, and $\gamma_{\text{red}, b}$ are the red noise parameters that are fit to the data set.
Lastly, the $\epsilon_{mb}$ covariance is taken to be infinite to remove any possible fitting degeneracy in the signal; this technical point is discussed in Ref.~\cite{Taylor:2021yjx}. 

Since we are not interested in the specific realization of the Gaussian noise but only in its statistical properties, we can analytically marginalize over all the possible noise realizations. We are therefore left with a marginalized likelihood that depends only on the covariance noise parameters discussed here and shown in Table~\ref{tab:parameters}. 
In matrix notation, this marginalized likelihood can be written as~\cite{vanHaasteren:2012hj,Lentati:2012xb,Taylor:2021yjx},
\begin{equation}
    p(\Vec{\delta t}|\Vec{\theta}) = \dfrac{\exp\left(-\displaystyle\frac{1}{2}\Vec{\Delta t}^T\mathbf{C}^{-1}\Vec{\Delta t}\right)}{\sqrt{\text{det}[2\pi\mathbf{C}]}} \, ,
    \label{eq:lkl_1}
\end{equation}
where the noise parameters are in $\vec{\theta}$, $\mathbf{C} = \mathbf{N} + \mathbf{TBT}^T$ is the total covariance matrix, $\vec{N}$ is the white noise covariance matrix which depends on $E_b$ and $Q_b$, $\mathbf{T} = [\mathbf{M}, \mathbf{F}]$ where $\mathbf{M}$ and $\mathbf{F}$ are the design and Fourier design matrices, respectively, and $\mathbf{B} = \mathrm{diag}(\mathbf{\infty}, \bm{\phi})$, where $\bm{\phi}$ is the red noise covariance matrix which depends on $\rho_\text{g}, A_\text{GWB}, \gamma_\text{GWB}, A_{\text{red}, b}$, and $\gamma_{\text{red}, b}$.

\begin{figure*}[ht!]
    \centering
    \includegraphics[width=0.75\linewidth]{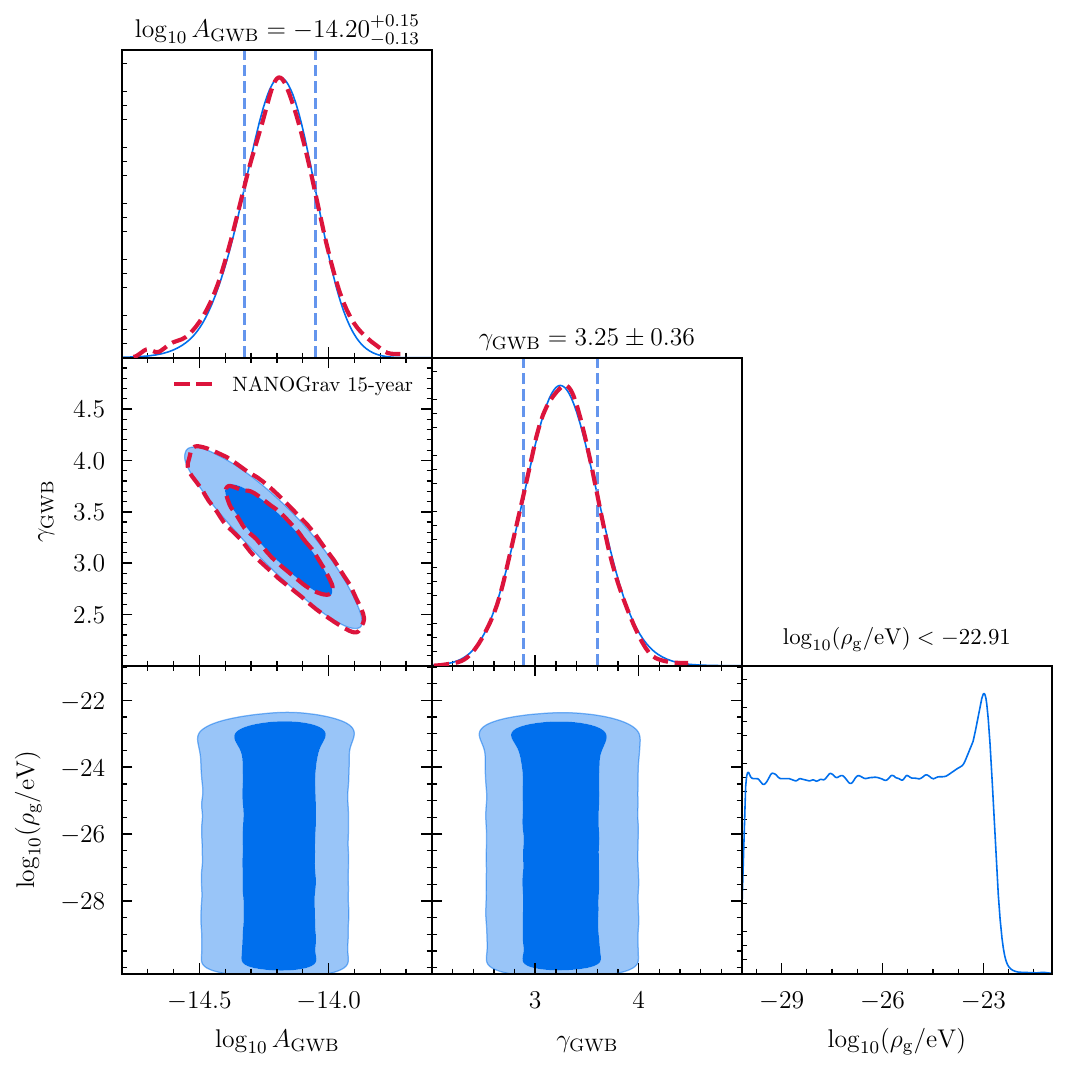}
    \caption{
    Marginalized posterior distributions for the GWB parameters and CSG spin scale, $\rhog$, following the analysis described in Sec.~\ref{sec:data_analysis}.
    In the top left and middle right panels, we report the 1d marginalized posterior distributions for the GWB parameters, along with $68\%$ credible intervals represented by dashed lines. The mean and $68\%$ credible interval edges for these distributions are summarized above the panels.
    The three panels in the lower left show the 2d posterior distributions, shading regions inside $68\%$ (dark blue) and $95\%$ (light blue) credible intervals.
    The bottom right panel is discussed in detail in Fig.~\ref{fig:1d_posterior_CSP}.
    The red dashed lines show the result obtained in the NANOGrav 15-year analysis~\cite{NANOGrav:2023gor}.
    } 
    \label{fig:sensitivity_figure}
\end{figure*}

\subsection{Bayesian Inference Pipeline}
\label{subsec:bayesian_inference}

With the marginalized likelihood discussed in Sec.~\ref{subsec:likelihood}, we can use Bayesian inference techniques to compute the posterior on the parameter of interest, $\rho_\text{g}$. 
The priors for the noise parameters and CSG spin scale are given in Table~\ref{tab:parameters}. The white noise, red noise, and SMBHB parameter priors are chosen to match previous searches for new physics in PTA datasets; see, for example, Ref.~\cite{NANOGrav:2023hvm}. 
Because the CSG signal depends on $\rho_\text{g} / \omega$ (Sec.~\ref{sec:theory}), the prior limits on $\rho_\text{g}$ were based on the minimal and maximal frequencies accessible by NANOGrav: $f_\text{min} \sim 1 / T$ and $f_\text{max} \sim 1 / t_\text{cad}$, where $t_\text{cad}$ is the measurement cadence. The lower prior limit on $\rho_\text{g}$ was chosen to be much smaller than $f_\text{min}$ (since the expectation is that the 95\% credible level should be near this), and the upper prior limit on $\rho_\text{g}$ is set to $1 / \text{week}$, which is approximately the scale of cadences in the NANOGrav 15-year data set~\cite{NANOGrav:2023gor}.

Given the priors and marginalized likelihood, posteriors can be computed. Computing the posteriors is computationally demanding, and we evaluate them with Monte Carlo sampling. Specifically, we use \textsf{PTArcade}, which uses the Markov chain Monte Carlo (MCMC) sampler implemented in the \textsf{PTMCMCSampler} package to sample parameter points, and also relies on \textsf{ENTERPRISE} and \textsf{ENTERPRISE\_EXTENSIONS}~\cite{enterprise, enterprise-ext}.
We make our \textsf{PTArcade} input and output files publicly available at Ref.~\cite{ambasht_2026_23254689}.

\section{Continuous Spin Graviton Spin Scale Sensitivity}
\label{sec:sensitivity}

\begin{figure}[ht!]
    \centering
    \includegraphics[width=\linewidth]{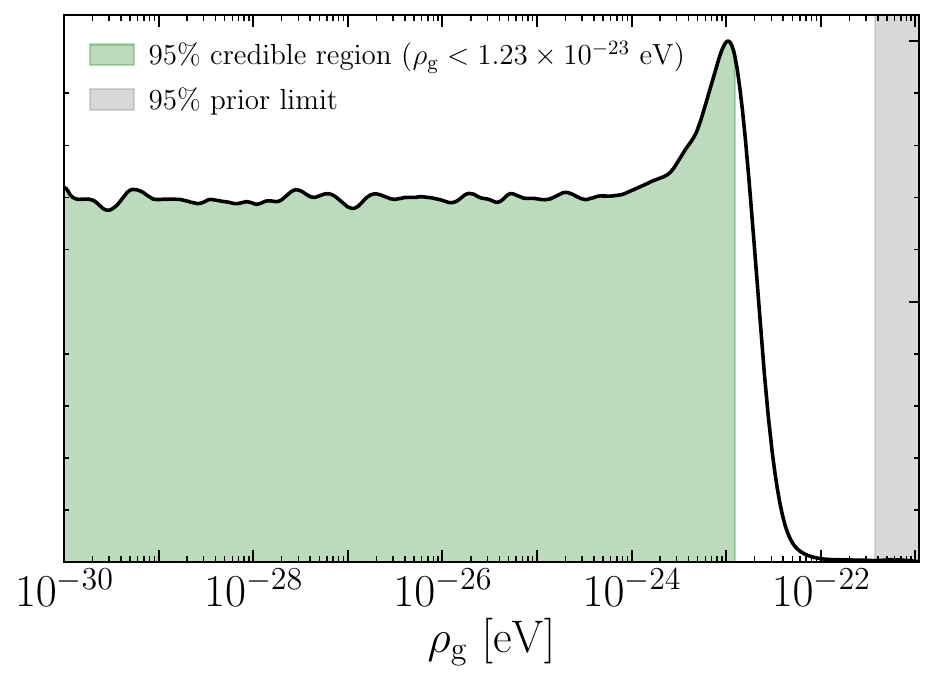}
    \caption{
    Marginalized posterior of the CSG spin scale, $\rhog$ (black).
    The green shaded region is the $95\%$ credible region, bounded to $\rhog < 1.23 \times 10^{-23} \, \text{eV}$.
    The grey shaded region denotes the $95^\text{th}$ percentile of the prior on $\rhog$.
    The plateau for small $\rhog$ corresponds to parameter space consistent with standard general relativity ($\rhog = 0$).
    }
    \label{fig:1d_posterior_CSP}
\end{figure}

Using the Bayesian analysis pipeline described in Sec.~\ref{sec:data_analysis}, we now demonstrate the sensitivity of the NANOGrav 15-year data set to $\rhog$.
The prior expectation for the result of this analysis is that $\rhog$ should be restricted to be much smaller than the smallest frequency in the data set, $\omega \sim 2 \pi / T \sim 10^{-23} \, \text{eV} \, (15 \, \text{yr} / T)$.
This is because when $\rhog / \omega \gg 1$, the angular and spectral properties of the ORF deviate significantly from the typical HD properties, as shown in Fig.~\ref{fig:modified_orf}.
The complete Bayesian analysis here confirms this preliminary expectation and explicitly quantifies the 95\% credible level limit on $\rhog$ in the NANOGrav 15-year data set.
The MCMC chains used $\mathcal{O}(10^6)$ points, reaching a Gelman-Rubin statistic of about $0.035$, representing extremely high convergence~\cite{gelman1992inference}. 

In Fig.~\ref{fig:sensitivity_figure} we show the marginalized posterior distributions on the GWB parameters ($A_\text{GWB}, \gamma_\text{GWB}$) and $\rhog$. 
We find that the 1d marginalized posteriors on $A_\text{GWB}$ and $\gamma_\text{GWB}$, along with the 2d marginalized posterior on $A_\text{GWB}, \gamma_\text{GWB}$ are consistent with the NANOGrav 15-year data set analysis (dashed red lines)~\cite{NANOGrav:2023gor}.
This is expected, since for small $\rhog$ the analysis performed here is identical to the standard GWB analysis, and larger $\rhog$ do not improve the fit enough to significantly shift the GWB parameter posteriors.
The 95\% credible level upper limit is derived from the 1d posterior distribution on $\rhog$, which is shown in the bottom right panel in Fig.~\ref{fig:sensitivity_figure}. 
We provide a detailed breakdown of this result in Fig.~\ref{fig:1d_posterior_CSP}.

In Fig.~\ref{fig:1d_posterior_CSP}, we present the 1d marginalized posterior distribution for $\rhog$. 
The plateau feature at small $\rhog$ corresponds to the region where the likelihood is agnostic to the exact value of $\rhog$. 
The peak at $\rhog \sim 10^{-23} \, \text{eV}$ marks the region where CSG-induced deviations slightly improve the fit to the NANOGrav 15-year data set, although the effect is not statistically significant.
Any further increase in $\rhog$ dampens the ORF, as shown in Fig.~\ref{fig:modified_orf}, causing the posterior to rapidly vanish.
We find that the upper limit on the CSG spin scale at the 95\% credible level from the NANOGrav 15-year data set is, 
\begin{align}
    \rho_\text{g} < 1.23 \times 10^{-23} \, \text{eV} \, .
    \label{eq:main_result_upper_bound}
\end{align}
This limit is independent of the upper bound on the prior range; the $95^\text{th}$ percentile of the $\rhog$ prior range is $1.36 \times 10^{-20} \, \mathrm{eV}$ (shaded gray), which is much larger than the limit in Eq.~\eqref{eq:main_result_upper_bound}.

\section{Conclusions}
\label{sec:conclusions}

Continuous spin particles are a theoretically self-consistent generalization of massless particles~\cite{Schuster:2023xqa}, possess a dimensionful spin scale $\rho$, and reduce to ordinary massless particles in the $\rho \rightarrow 0$ limit.
Recent work demonstrated that a non-zero graviton spin scale, $\rhog$, modifies the GWB-induced timing delays in PTAs compared to the ordinary $\rhog \rightarrow 0$ limit~\cite{Kundu:2025fsd}.
Therefore, since PTAs have seen evidence for a GWB consistent with $\rhog = 0$~\cite{Xu:2023wog,EPTA:2023fyk,NANOGrav:2023gor,Reardon:2023gzh}, they can place tight constraints on $\rhog$.
In this paper we used the NANOGrav 15-year data set~\cite{NANOGrav:2023hde} to place an upper limit at the $95\%$ credible level of $\rhog < 1.23 \times 10^{-23} \, \text{eV}$.

In Sec.~\ref{sec:theory} we derived the CSG-modified timing delay relevant for PTAs, building on recent work in Ref.~\cite{Kundu:2025fsd}.
From this CSG-induced timing delay we derived a modified ORF (Sec.~\ref{subsec:modified_orf}), which we found to have suppressed angular correlations, relative to the $\rhog \rightarrow 0$ limit, when $\rhog / \omega \gg 1$ (Fig.~\ref{fig:modified_orf}), where $\omega$ is the GW frequency.
We implemented this modified ORF into the PTA-standard Bayesian analysis pipeline (Sec.~\ref{sec:data_analysis}), performed parameter inference using the NANOGrav 15-year data set (Sec.~\ref{sec:sensitivity}), and show the 1d marginalized posterior on $\rhog$ in Fig.~\ref{fig:1d_posterior_CSP}.

Throughout we have assumed that the SGWB production is unaffected by a finite $\rhog$.
We expect this to be a conservative assumption, since modified SGWB production would lead to additional phenomenology that would help distinguish the $\rhog \neq 0$ and $\rhog = 0$ limits.
For instance, if CSG modes with helicity $\neq \pm 2$ are emitted during production, a suppressed contribution of $\mathcal{O}(\rhog v / \omega)$, where $v$ is the velocity of the merging supermassive black holes~\cite{Schuster:2023xqa,Kundu:2025fsd}, they would additionally alter the timing residual at $\mathcal{O}(\rhog / \omega)$, resulting in a total effect on the timing residual at $\mathcal{O}(\rhog^2 v / \omega^2)$.
While expected to be suppressed by $v$, such an effect could imprint unique, identifiable features in the modified ORF.
Furthermore, if the helicity $\pm 2$ mode production is suppressed by $\rhog / \omega$, this would effectively appear as an additional suppression to the modified ORF, i.e., an enhanced turnover in the right panel of Fig.~\ref{fig:modified_orf}.
However, since the production frequency of the GWs is higher than the detected frequency due to redshifting, suppression at production would appear at lower observed frequencies than the effects discussed here.

Finally, the upper limit on $\rhog$ could be further improved by applying the framework detailed here to future PTA datasets, e.g., the upcoming International Pulsar Timing Array data sets~\cite{InternationalPulsarTimingArray:2023mzf}, with the expectation that the longest data set with evidence for a GWB will provide the most stringent limits.
Additional future work could study the implications a CSG would have on non-SMBHB produced GWBs, for example those produced cosmologically by physics beyond the Standard Model~\cite{Vilenkin:1984ib,Guzzetti:2016mkm,Caprini:2019egz,Domenech:2021ztg,NANOGrav:2023hvm}.
Some of these models predict enhanced low-frequency GW production, yielding timing residuals too large to be consistent with PTA datasets. However, a CSG might be able to reconcile this tension, since a non-zero $\rhog$ suppresses the PTA signal.

\acknowledgments
We would like to thank Shayarneel Kundu, Rafael R. Lino dos Santos, and Philip Schuster for helpful discussion.
TT is supported by the DOE grant DE-SC0015655.
This work made use of the Illinois Campus Cluster, a computing resource that is operated by the Illinois Campus Cluster Program (ICCP) in conjunction with the National Center for Supercomputing Applications (NCSA) and which is supported by funds from the University of Illinois Urbana-Champaign. 

\appendix

\section{Numerical Evaluation of the Modified Overlap Reduction Function}
\label{app:modified_orf_eval}

The modified ORF in Eq.~\eqref{eq:money_ORF} is calculated by numerically performing the integral over $\hat{\mathbf{\Omega}}$ in the computational frame, i.e., with $\hat{\mathbf{d}}_a, \hat{\mathbf{d}}_b$ given by Eq.~\eqref{eq:computational_frame}.
Because the modified ORF depends solely on the angle between pulsars and $\rhog / \omega$, we avoid evaluating Eq.~\eqref{eq:money_ORF} for every pair by precomputing $\Gamma(\xi_{ab})$ on a grid and interpolating as needed.

The grid in angular separation is 500 linearly spaced points between $10^{-3} \leq \xi_{ab} \leq \pi$.
The grid in $\rhog / \omega$ is built from three different grids.
First, a linearly spaced grid of $10^3$ points between $0 \leq \rhog / \omega \leq 10^{-1}$ is used to finely cover the range of small $\rho_\text{g} / \omega$.
This is effective in capturing small departures from the standard HD ORF.
Second, a logarithmically spaced grid of $7 \times 10^3$ points between $10^{-1} \leq \rhog / \omega \leq 10^3$ is used to capture the region with maximal Bessel function oscillations.
Third, we use a logarithmically spaced grid of $2 \times 10^3$ points between $10^3 \leq \rhog / \omega \leq 10^4$.
This grid captures the large $\rhog / \omega$ suppression of the ORF. For $\rho_\text{g} / \omega > 10^4$, the ORF is clipped to the value obtained from $\rho_\text{g} / \omega = 10^4$.
The absolute and relative errors of this look-up grid for the modified ORF are $\mathcal{O}(10^{-6})$, leading to negligible interpolation error.

\bibliographystyle{utphys3}
\bibliography{references}

\onecolumngrid

\section{NANOGrav Collaboration Author List}
\label{app:complete_author_list}
\begin{itemize}
\item Gabriella Agazie\;\orcidlink{0000-0001-5134-3925}\\\textit{Center for Gravitation, Cosmology and Astrophysics, Department of Physics and Astronomy, University of Wisconsin-Milwaukee, P.O. Box 413, Milwaukee, WI 53201, USA}
\item Nikita Agarwal\;\orcidlink{0000-0001-7544-7876}\\\textit{Department of Physics and Astronomy, West Virginia University, P.O. Box 6315, Morgantown, WV 26506, USA}\\\textit{Center for Gravitational Waves and Cosmology, West Virginia University, Chestnut Ridge Research Building, Morgantown, WV 26505, USA}
\item Anjana Ashok\;\orcidlink{0000-0002-8395-957X}\\\textit{Department of Physics, Oregon State University, Corvallis, OR 97331, USA}
\item Jeremy G. Baier\;\orcidlink{0000-0002-4972-1525}\\\textit{Department of Physics, Oregon State University, Corvallis, OR 97331, USA}
\item Paul T. Baker\;\orcidlink{0000-0003-2745-753X}\\\textit{Department of Physics and Astronomy, Widener University, One University Place, Chester, PA 19013, USA}
\item Bence B\'ecsy\;\orcidlink{0000-0003-0909-5563}\\\textit{Institute for Gravitational Wave Astronomy and School of Physics and Astronomy, University of Birmingham, Edgbaston, Birmingham B15 2TT, UK}
\item Laura Blecha\;\orcidlink{0000-0002-2183-1087}\\\textit{Physics Department, University of Florida, Gainesville, FL 32611, USA}
\item Adam Brazier\;\orcidlink{0000-0001-6341-7178}\\\textit{Cornell Center for Astrophysics and Planetary Science and Department of Astronomy, Cornell University, Ithaca, NY 14853, USA}\\\textit{Cornell Center for Advanced Computing, Cornell University, Ithaca, NY 14853, USA}
\item Paul R. Brook\;\orcidlink{0000-0003-3053-6538}\\\textit{Institute for Gravitational Wave Astronomy and School of Physics and Astronomy, University of Birmingham, Edgbaston, Birmingham B15 2TT, UK}
\item Sarah Burke-Spolaor\;\orcidlink{0000-0003-4052-7838}\\\textit{Department of Physics and Astronomy, West Virginia University, P.O. Box 6315, Morgantown, WV 26506, USA}\\\textit{Center for Gravitational Waves and Cosmology, West Virginia University, Chestnut Ridge Research Building, Morgantown, WV 26505, USA}
\item Rand Burnette\;\orcidlink{0009-0008-3649-0618}\\\textit{Department of Physics, Oregon State University, Corvallis, OR 97331, USA}
\item Robin Case\;\orcidlink{0009-0007-4346-8921}\\\textit{Department of Physics, Oregon State University, Corvallis, OR 97331, USA}
\item J. Andrew Casey-Clyde\;\orcidlink{0000-0002-5557-4007}\\\textit{Department of Physics, University of Connecticut, 196 Auditorium Road, U-3046, Storrs, CT 06269-3046, USA}
\item Maria Charisi\;\orcidlink{0000-0003-3579-2522}\\\textit{Department of Physics and Astronomy, Washington State University, Pullman, WA 99163, USA}\\\textit{Institute of Astrophysics, FORTH, GR-71110, Heraklion, Greece}
\item Shami Chatterjee\;\orcidlink{0000-0002-2878-1502}\\\textit{Cornell Center for Astrophysics and Planetary Science and Department of Astronomy, Cornell University, Ithaca, NY 14853, USA}
\item Tyler Cohen\;\orcidlink{0000-0001-7587-5483}\\\textit{Department of Physics, New Mexico Institute of Mining and Technology, 801 Leroy Place, Socorro, NM 87801, USA}
\item James M. Cordes\;\orcidlink{0000-0002-4049-1882}\\\textit{Cornell Center for Astrophysics and Planetary Science and Department of Astronomy, Cornell University, Ithaca, NY 14853, USA}
\item Neil J. Cornish\;\orcidlink{0000-0002-7435-0869}\\\textit{Department of Physics, Montana State University, Bozeman, MT 59717, USA}
\item Fronefield Crawford\;\orcidlink{0000-0002-2578-0360}\\\textit{Department of Physics and Astronomy, Franklin \& Marshall College, P.O. Box 3003, Lancaster, PA 17604, USA}
\item H. Thankful Cromartie\;\orcidlink{0000-0002-6039-692X}\\\textit{Department of Physics and Astronomy, Vanderbilt University, 2301 Vanderbilt Place, Nashville, TN 37235, USA}
\item Megan E. DeCesar\;\orcidlink{0000-0002-2185-1790}\\\textit{Department of Physics and Astronomy, George Mason University, Fairfax, VA 22030}
\item Paul B. Demorest\;\orcidlink{0000-0002-6664-965X}\\\textit{National Radio Astronomy Observatory, 1003 Lopezville Rd., Socorro, NM 87801, USA}
\item Heling Deng\;\orcidlink{0000-0002-1918-5477}\\\textit{Columbia Astrophysics Laboratory, Columbia University, 538 West 120th Street, New York, NY 10027, USA}
\item Lankeswar Dey\;\orcidlink{0000-0002-2554-0674}\\\textit{Institute of Astrophysics, FORTH, GR-71110, Heraklion, Greece}
\item Timothy Dolch\;\orcidlink{0000-0001-8885-6388}\\\textit{Department of Physics and Astronomy, University of New Mexico, Albuquerque, NM 87131, USA}\\\textit{Department of Physics, Hillsdale College, 33 E. College Street, Hillsdale, MI 49242, USA}\\\textit{Eureka Scientific, 2452 Delmer Street, Suite 100, Oakland, CA 94602-3017, USA}\\\textit{SETI Institute, 339 N Bernardo Ave Suite 200, Mountain View, CA 94043, USA}
\item Graham M. Doskoch\;\orcidlink{0000-0002-4219-6908}\\\textit{Department of Physics and Astronomy, West Virginia University, P.O. Box 6315, Morgantown, WV 26506, USA}\\\textit{Center for Gravitational Waves and Cosmology, West Virginia University, Chestnut Ridge Research Building, Morgantown, WV 26505, USA}
\item William Fiore\;\orcidlink{0000-0001-5645-5336}\\\textit{Department of Physics and Astronomy, University of British Columbia, 6224 Agricultural Road, Vancouver, BC V6T 1Z1, Canada}
\item Emmanuel Fonseca\;\orcidlink{0000-0001-8384-5049}\\\textit{Department of Physics and Astronomy, West Virginia University, P.O. Box 6315, Morgantown, WV 26506, USA}\\\textit{Center for Gravitational Waves and Cosmology, West Virginia University, Chestnut Ridge Research Building, Morgantown, WV 26505, USA}
\item Gabriel E. Freedman\;\orcidlink{0000-0001-7624-4616}\\\textit{NASA Goddard Space Flight Center, Greenbelt, MD 20771, USA}
\item Emiko C. Gardiner\;\orcidlink{0000-0002-8857-613X}\\\textit{Department of Astronomy, University of California, Berkeley, 501 Campbell Hall \#3411, Berkeley, CA 94720, USA}
\item Kyle A. Gersbach\;\orcidlink{0009-0009-5393-0141}\\\textit{Department of Physics and Astronomy, Vanderbilt University, 2301 Vanderbilt Place, Nashville, TN 37235, USA}
\item Joseph Glaser\;\orcidlink{0000-0003-4090-9780}\\\textit{Department of Physics and Astronomy, West Virginia University, P.O. Box 6315, Morgantown, WV 26506, USA}\\\textit{Center for Gravitational Waves and Cosmology, West Virginia University, Chestnut Ridge Research Building, Morgantown, WV 26505, USA}
\item Deborah C. Good\;\orcidlink{0000-0003-1884-348X}\\\textit{Department of Physics and Astronomy, University of Montana, 32 Campus Drive, Missoula, MT 59812}
\item Kayhan G\"ultekin\;\orcidlink{0000-0002-1146-0198}\\\textit{Department of Astronomy and Astrophysics, University of Michigan, Ann Arbor, MI 48109, USA}
\item Aiden Gundersen\;\orcidlink{0009-0004-2085-6348}\\\textit{Department of Physics, Montana State University, Bozeman, MT 59717, USA}
\item C. J. Harris\;\orcidlink{0000-0002-4231-7802}\\\textit{Department of Astronomy and Astrophysics, University of Michigan, Ann Arbor, MI 48109, USA}
\item Jeffrey S. Hazboun\;\orcidlink{0000-0003-2742-3321}\\\textit{Department of Physics, Oregon State University, Corvallis, OR 97331, USA}
\item Ross J. Jennings\;\orcidlink{0000-0003-1082-2342}\\\textit{Department of Physics and Astronomy, West Virginia University, P.O. Box 6315, Morgantown, WV 26506, USA}\\\textit{Center for Gravitational Waves and Cosmology, West Virginia University, Chestnut Ridge Research Building, Morgantown, WV 26505, USA}
\item Aaron D. Johnson\;\orcidlink{0000-0002-7445-8423}\\\textit{Center for Gravitation, Cosmology and Astrophysics, Department of Physics and Astronomy, University of Wisconsin-Milwaukee, P.O. Box 413, Milwaukee, WI 53201, USA}\\\textit{Division of Physics, Mathematics, and Astronomy, California Institute of Technology, Pasadena, CA 91125, USA}
\item David L. Kaplan\;\orcidlink{0000-0001-6295-2881}\\\textit{Center for Gravitation, Cosmology and Astrophysics, Department of Physics and Astronomy, University of Wisconsin-Milwaukee, P.O. Box 413, Milwaukee, WI 53201, USA}
\item Anala K. Sreekumar\;\orcidlink{0009-0001-7906-8520}\\\textit{Department of Physics and Astronomy, West Virginia University, P.O. Box 6315, Morgantown, WV 26506, USA}\\\textit{Center for Gravitational Waves and Cosmology, West Virginia University, Chestnut Ridge Research Building, Morgantown, WV 26505, USA}
\item Luke Zoltan Kelley\;\orcidlink{0000-0002-6625-6450}\\\textit{Astrophysics Working Group, NANOGrav Collaboration, Berkeley, CA, USA}
\item Joey S. Key\;\orcidlink{0000-0003-0123-7600}\\\textit{University of Washington Bothell, 18115 Campus Way NE, Bothell, WA 98011, USA}
\item Nima Laal\;\orcidlink{0000-0002-9197-7604}\\\textit{Department of Physics and Astronomy, Vanderbilt University, 2301 Vanderbilt Place, Nashville, TN 37235, USA}
\item Michael T. Lam\;\orcidlink{0000-0003-0721-651X}\\\textit{SETI Institute, 339 N Bernardo Ave Suite 200, Mountain View, CA 94043, USA}
\item William G. Lamb\;\orcidlink{0000-0003-1096-4156}\\\textit{Department of Physics and Astronomy, Vanderbilt University, 2301 Vanderbilt Place, Nashville, TN 37235, USA}
\item Bjorn Larsen\;\orcidlink{0000-0001-6436-8216}\\\textit{Department of Physics, Yale University, New Haven, CT 06511, USA}
\item T. Joseph W. Lazio\;\orcidlink{0009-0003-8984-388X}\\\textit{Jet Propulsion Laboratory, California Institute of Technology, 4800 Oak Grove Drive, Pasadena, CA 91109, USA}
\item Natalia Lewandowska\;\orcidlink{0000-0003-0771-6581}\\\textit{Department of Physics and Astronomy, State University of New York at Oswego, Oswego, NY 13126, USA}
\item Tingting Liu\;\orcidlink{0000-0001-5766-4287}\\\textit{Department of Physics and Astronomy, Georgia State University, 25 Park Place, Suite 605, Atlanta, GA 30303, USA}
\item Ryan S. Lynch\;\orcidlink{0000-0001-5229-7430}\\\textit{Green Bank Observatory, P.O. Box 2, Green Bank, WV 24944, USA}
\item Chung-Pei Ma\;\orcidlink{0000-0002-4430-102X}\\\textit{Department of Astronomy, University of California, Berkeley, 501 Campbell Hall \#3411, Berkeley, CA 94720, USA}\\\textit{Department of Physics, University of California, Berkeley, CA 94720, USA}
\item Dustin R. Madison\;\orcidlink{0000-0003-2285-0404}\\\textit{Department of Physics, Occidental College, 1600 Campus Road, Los Angeles, CA 90041, USA}
\item Ashley Martsen\;\orcidlink{0000-0001-8313-0895}\\\textit{Department of Physics and Astronomy, West Virginia University, P.O. Box 6315, Morgantown, WV 26506, USA}\\\textit{Center for Gravitational Waves and Cosmology, West Virginia University, Chestnut Ridge Research Building, Morgantown, WV 26505, USA}
\item Cayenne Matt\;\orcidlink{0000-0002-9710-6527}\\\textit{Department of Astronomy and Astrophysics, University of Michigan, Ann Arbor, MI 48109, USA}
\item James W. McKee\;\orcidlink{0000-0002-2885-8485}\\\textit{Department of Physics and Astronomy, Union College, Schenectady, NY 12308, USA}
\item Maura A. McLaughlin\;\orcidlink{0000-0001-7697-7422}\\\textit{Department of Physics and Astronomy, West Virginia University, P.O. Box 6315, Morgantown, WV 26506, USA}\\\textit{Center for Gravitational Waves and Cosmology, West Virginia University, Chestnut Ridge Research Building, Morgantown, WV 26505, USA}
\item Patrick M. Meyers\;\orcidlink{0000-0002-2689-0190}\\\textit{ETH Zurich, Institute for Particle Physics and Astrophysics, Wolfgang-Pauli-Strasse 27, 8093 Zurich, Switzerland}
\item Matthew T. Miles\;\orcidlink{0000-0002-5455-3474}\\\textit{Department of Physics and Astronomy, Vanderbilt University, 2301 Vanderbilt Place, Nashville, TN 37235, USA}
\item Chiara M. F. Mingarelli\;\orcidlink{0000-0002-4307-1322}\\\textit{Department of Physics, Yale University, New Haven, CT 06511, USA}\\\textit{Center for Computational Astrophysics, Flatiron Institute, 162 5th Avenue, New York, NY 10010, USA}
\item Andrea Mitridate\;\orcidlink{0000-0003-2898-5844}\\\textit{Abdus Salam Centre for Theoretical Physics, Imperial College London, London SW7 2AZ, UK}
\item David J. Nice\;\orcidlink{0000-0002-6709-2566}\\\textit{Department of Physics, Lafayette College, Easton, PA 18042, USA}
\item Shania A. Nichols\;\orcidlink{0009-0001-1750-3531}\\\textit{SETI Institute, 339 N Bernardo Ave Suite 200, Mountain View, CA 94043, USA}
\item Stella Koch Ocker\;\orcidlink{0000-0002-4941-5333}\\\textit{Division of Physics, Mathematics, and Astronomy, California Institute of Technology, Pasadena, CA 91125, USA}\\\textit{The Observatories of the Carnegie Institution for Science, Pasadena, CA 91101, USA}
\item Daniel J. Oliver\;\orcidlink{0000-0002-7374-6925}\\\textit{Department of Physics, Oregon State University, Corvallis, OR 97331, USA}
\item Ken D. Olum\;\orcidlink{0000-0002-2027-3714}\\\textit{Institute of Cosmology, Department of Physics and Astronomy, Tufts University, Medford, MA 02155, USA}
\item Polina Petrov\;\orcidlink{0000-0001-5681-4319}\\\textit{Department of Physics and Astronomy, Vanderbilt University, 2301 Vanderbilt Place, Nashville, TN 37235, USA}
\item Nihan S. Pol\;\orcidlink{0000-0002-8826-1285}\\\textit{Department of Physics, Texas Tech University, Box 41051, Lubbock, TX 79409, USA}
\item Scott M. Ransom\;\orcidlink{0000-0001-5799-9714}\\\textit{National Radio Astronomy Observatory, 520 Edgemont Road, Charlottesville, VA 22903, USA}
\item Paul S. Ray\;\orcidlink{0000-0002-5297-5278}\\\textit{Space Science Division, Naval Research Laboratory, Washington, DC 20375-5352, USA}
\item Joseph D. Romano\;\orcidlink{0000-0003-4915-3246}\\\textit{Department of Physics, Texas Tech University, Box 41051, Lubbock, TX 79409, USA}
\item Jessie C. Runnoe\;\orcidlink{0000-0001-8557-2822}\\\textit{Department of Physics and Astronomy, Vanderbilt University, 2301 Vanderbilt Place, Nashville, TN 37235, USA}
\item Alexander Saffer\;\orcidlink{0000-0001-7832-9066}\\\textit{National Radio Astronomy Observatory, 520 Edgemont Road, Charlottesville, VA 22903, USA}
\item Shashwat C. Sardesai\;\orcidlink{0009-0006-5476-3603}\\\textit{Center for Gravitation, Cosmology and Astrophysics, Department of Physics and Astronomy, University of Wisconsin-Milwaukee, P.O. Box 413, Milwaukee, WI 53201, USA}
\item Kai Schmitz\;\orcidlink{0000-0003-2807-6472}\\\textit{Institute for Theoretical Physics, University of Münster, 48149 Münster, Germany}
\item Levi Schult\;\orcidlink{0000-0001-6425-7807}\\\textit{Department of Physics and Astronomy, Vanderbilt University, 2301 Vanderbilt Place, Nashville, TN 37235, USA}
\item Xavier Siemens\;\orcidlink{0000-0002-7778-2990}\\\textit{Department of Physics, Oregon State University, Corvallis, OR 97331, USA}\\\textit{Center for Gravitation, Cosmology and Astrophysics, Department of Physics and Astronomy, University of Wisconsin-Milwaukee, P.O. Box 413, Milwaukee, WI 53201, USA}
\item Sophia V. Sosa Fiscella\;\orcidlink{0000-0002-5176-2924}\\\textit{ASTRON, Netherlands Institute for Radio Astronomy, Oude Hoogeveensedijk 4, 7991 PD Dwingeloo, The Netherlands}
\item Ingrid H. Stairs\;\orcidlink{0000-0001-9784-8670}\\\textit{Department of Physics and Astronomy, University of British Columbia, 6224 Agricultural Road, Vancouver, BC V6T 1Z1, Canada}
\item Daniel R. Stinebring\;\orcidlink{0000-0002-1797-3277}\\\textit{Department of Physics and Astronomy, Oberlin College, Oberlin, OH 44074, USA}
\item Jacob Taylor\;\orcidlink{0000-0001-9118-5589}\\\textit{Department of Physics, Oregon State University, Corvallis, OR 97331, USA}
\item Stephen R. Taylor\;\orcidlink{0000-0003-0264-1453}\\\textit{Department of Physics and Astronomy, Vanderbilt University, 2301 Vanderbilt Place, Nashville, TN 37235, USA}
\item Mercedes S. Thompson\;\orcidlink{0009-0001-5938-5000}\\\textit{Department of Physics and Astronomy, University of British Columbia, 6224 Agricultural Road, Vancouver, BC V6T 1Z1, Canada}
\item Jacob E. Turner\;\orcidlink{0000-0002-2451-7288}\\\textit{Green Bank Observatory, P.O. Box 2, Green Bank, WV 24944, USA}
\item Michele Vallisneri\;\orcidlink{0000-0002-4162-0033}\\\textit{ETH Zurich, Institute for Particle Physics and Astrophysics, Wolfgang-Pauli-Strasse 27, 8093 Zurich, Switzerland}
\item Rutger van~Haasteren\;\orcidlink{0000-0002-6428-2620}\\\textit{Max-Planck-Institut f{\"u}r Gravitationsphysik (Albert-Einstein-Institut), Callinstra{\ss}e 38, D-30167 Hannover, Germany}\\\textit{Leibniz Universit{\"a}t Hannover, D-30167 Hannover, Germany}
\item Sarah J. Vigeland\;\orcidlink{0000-0003-4700-9072}\\\textit{Center for Gravitation, Cosmology and Astrophysics, Department of Physics and Astronomy, University of Wisconsin-Milwaukee, P.O. Box 413, Milwaukee, WI 53201, USA}
\item Kevin P. Wilson\;\orcidlink{0000-0003-4231-2822}\\\textit{Department of Physics and Astronomy, West Virginia University, P.O. Box 6315, Morgantown, WV 26506, USA}\\\textit{Center for Gravitational Waves and Cosmology, West Virginia University, Chestnut Ridge Research Building, Morgantown, WV 26505, USA}
\item Caitlin A. Witt\;\orcidlink{0000-0002-6020-9274}\\\textit{Department of Physics, Wake Forest University, 1834 Wake Forest Road, Winston-Salem, NC 27109}
\item David Wright\;\orcidlink{0000-0003-1562-4679}\\\textit{Department of Physics, Oregon State University, Corvallis, OR 97331, USA}
\item Olivia Young\;\orcidlink{0000-0002-0883-0688}\\\textit{School of Physics and Astronomy, Rochester Institute of Technology, Rochester, NY 14623, USA}\\\textit{Laboratory for Multiwavelength Astrophysics, Rochester Institute of Technology, Rochester, NY 14623, USA}
\end{itemize}

\end{document}